\documentclass[10pt, a4paper]{article}
\usepackage{color,graphicx}
\usepackage{amsmath}

\usepackage{endnotes} 

\usepackage{html}  

\usepackage{url}	 

\newcommand{\beq}{\begin{equation}}
  \newcommand{\eeq}{\end{equation}}
  \newcommand{\beqno}{\begin{displaymath}}
  \newcommand{\eeqno}{\end{displaymath}}
  \newcommand{\beqar}{\begin{eqnarray}}
  \newcommand{\eeqar}{\end{eqnarray}}
  \newcommand{\beqarno}{\begin{eqnarray*}}
  \newcommand{\eeqarno}{\end{eqnarray*}}

\newcommand{\cref}[1]{(\ref{#1})}

\newcommand{\ba}{\begin{array}}
\newcommand{\ea}{\end{array}}

\newcommand{\ben}{\begin{enumerate}}
  \newcommand{\en}{ \end{enumerate}}

\newcommand{\bei}{\begin{itemize}}
  \newcommand{\ei}{ \end{itemize}}

\newcommand{\bed}{\begin{description}}
  \newcommand{\ed}{\end{description}}

\newcommand{\bec}{\begin{center}}
  \newcommand{\ec}{\end{center}}

\newcommand{\bprop}{\begin{proposition}}
  \newcommand{\eprop}{\end{proposition}}

\newcommand{\bdf}{\begin{definition}}
  \newcommand{\edf}{\end{definition}}

\newcommand{\bth}{\begin{theorem}}
  \newcommand{\eth}{\end{theorem}}

\newcommand{\bcon}{\begin{con}}
  \newcommand{\econ}{\end{con}}

\newcommand{\bcor}{\begin{corollary}}
  \newcommand{\ecor}{\end{corollary}}

\newcommand{\bpr}{\begin{problem}}
  \newcommand{\epr}{\end{problem}}

\newcommand{\blem}{\begin{lemma}}
  \newcommand{\elem}{\end{lemma}}

\newcommand{\brem}{\begin{remark}}
  \newcommand{\erem}{\end{remark}}
 
\newcommand{\bass}{\begin{assumption}}
  \newcommand{\eass}{\end{assumption}}

\newcommand{\bres}{\begin{result}}
  \newcommand{\eres}{\end{result}}

\newcommand{\bexm}{\begin{example}}
  \newcommand{\eexm}{\end{example}}
  
  \newcommand{\bema}{\begin{main}}
  \newcommand{\ema}{\end{main}}

\newtheorem{theorem}{Theorem}[section]
  \newtheorem{assumption}[theorem]{Assumption}
  \newtheorem{definition}[theorem]{Definition}
  \newtheorem{corollary}[theorem]{Corollary}
  \newtheorem{lemma}[theorem]{Lemma}
  \newtheorem{proposition}[theorem]{Proposition}
 \newtheorem{result}[theorem]{Result}
  \newtheorem{example}[theorem]{Example}
  \newtheorem{"definition"}[theorem]{"Definition"}
  \newtheorem{remark}[theorem]{Remark}
  \newtheorem{con}[theorem]{Conjecture}
  \newtheorem{problem}[theorem]{Problem}
 \newtheorem{main}[theorem]{Main question}

\def\proof{\bigskip \penalty 25\noindent{\bf Proof. }}

\def\endproof{\blackslug \bigskip}

\def\blackslug{\hbox{\hskip 1pt \vrule width 4pt height 6pt depth 1.5pt
  \hskip 1pt}}

\newcommand{\krull}[1]{\left\{ {#1} \right\}}

\newcommand{\bracket}[1]{\left[ {#1} \right]}

\newcommand{\pa}[1]{\left( {#1} \right)}

\newcommand{\model}{(\Omega,{\cal F},\Pi,X,\underline{\cal F})}

\usepackage{amsfonts}

\begin{document}

\title{An Explicit Formula for Optimal Portfolios in Complete Wiener Driven Markets: a Functional It\^o Calculus Approach}

\author{Kristoffer Lindensj\"o
\\ 
							\small{Department of Mathematics}\\
							\small{Stockholm University}\\
							\small{SE-106 91 Stockholm,}
							\small{Sweden}\\
              \small{kristoffer.lindensjo@math.su.se}\\
              \small{+46 70 444 10 07}
}   

\date{\today}

\maketitle

\begin{abstract}
\noindent We consider a standard optimal investment problem in a complete financial market driven by a Wiener process and derive an explicit formula for the optimal portfolio process in terms of the vertical derivative from functional It\^o calculus. An advantage with this approach compared to the Malliavin calculus approach is that it relies only on an integrability condition. \\

{\small \noindent {\bf AMS MSC2010:} 91G10;  93E20; 97M30; 91G80\\

}

\end{abstract}

\section{Introduction}
Optimal investment and consumption problems are among the most important problems in mathematical finance. Problems of this type were first studied in a Markovian framework using standard stochastic control methods, see e.g. \cite{merton69,merton71}. The martingale method studied in the present paper was developed in e.g. \cite{karatzas87,pliska86}. 

We consider the optimal investment problem of maximizing the expected value of a general utility function of terminal wealth in a standard complete financial market driven by a Wiener process, see Section \ref{Setup} and Section \ref{sec:optimal-port} for a detailed description of the problem. It is well-known that if $X^*$ is the optimal wealth process and $H$ is the state price density, then the discounted wealth process, given by $X^*(t)H(t)$, is a martingale, and it is therefore possible to implicitly characterize the optimal portfolio $\pi^*$ using the standard martingale representation theorem. In the present paper, we use this result and a constructive martingale representation theorem from functional It\^o calculus to derive an explicit formula for the optimal portfolio $\pi^*$ in terms of the \emph{vertical derivative}, see Theorem \ref{mainres1}.

There is a large literature on optimal investment problems using Malliavin calculus and in particular the Clark-Ocone formula. 
Using this approach it is possible to derive explicit formulas for optimal portfolios in terms of Malliavin derivatives. One of the first papers in this direction was \cite{oconeKaratzas}. Other papers using the Malliavin calculus approach to optimal investment and consumption problems include \cite{benth2003explicit,detemple2005closed,di2009optimal,lakner1998optimal,lakner2006portfolio,pham2001optimal,putschogl2008optimal}.

An advantage with the functional It\^o calculus approach to optimal portfolios proposed in the present paper compared to the Malliavin calculus approach is that it only relies on an integrability condition, whereas the Malliavin calculus approach relies on a differentiability condition in the Malliavin sense, which requires further restrictions on the financial market. This point is elaborated in Section \ref{mall-comp}. 

Additionally, a purpose of the present paper is to point out an area of application in which the theory of functional It\^o calculus can be applied.

The structure of the present paper is as follows. Section \ref{preliminaries} contains a non-technical account of the relevant parts of functional It\^o calculus. Section \ref{Setup} contains a description of the financial market. Section \ref{sec:optimal-port} presents the optimal investment problem and the explicit formula for the optimal portfolio. In Section \ref{mall-comp} we describe in what sense the approach of the present paper requires less restrictions on the financial market compared  to the Malliavin calculus approach. In section \ref{examples} we illustrate the explicit formula for  optimal portfolios by studying two well-known examples.

\brem {\normalfont A method for the computation of explicit approximations to functional It\^o calculus martingale representations is studied in \cite{cont2016weak}. Numerical studies of optimal portfolios using the Malliavin calculus approach can be found in \cite{detemple2003monte,takahashi2004asymptotic}.}
\erem

\brem {\normalfont The functional It\^o formula is in \cite{pang2015application} used in the study of an optimal investment problem where asset prices are modeled by a particular stochastic delay differential equation. Moreover, the particular structure of the problem implies that an HJB equation for an optimal value function depending only on a finite number (four) of state variables can be derived using the functional It\^o formula, and a corresponding verification theorem is proved. The approach is therefore fundamentally different from that of the present paper.}
\erem

\section{Martingale representation in functional It\^o calculus} \label{preliminaries}

Functional It\^o calculus was proposed in \cite{dupire}. It was developed into a coherent theory in e.g. \cite{ContBook,contpathspace,cont10,cont2013,cont2016weak}. This section contains a non-technical account of the relevant parts of functional It\^o calculus. For a comprehensive account we refer to \cite{ContBook,cont2013}.

Consider the space of c\`adl\`ag paths $\Omega = D([0,T],\mathbb{R}^n)$, where $T<\infty$. 
The value of a path $\omega$ at a fixed $t$ is denoted by $\omega(t)$ and a path stopped at $t$ is denoted by $\omega_t$, i.e. $\omega_{t}(s) = \omega(t \wedge s), 0 \leq s \leq T$. Let $F:[0,T]\times\Omega\rightarrow \mathbb{R}$ be a functional of paths that is non-anticipative in the sense that $F(t,\omega_t) = F(t,\omega)$. 

The horizontal derivative of $F$ is defined by
\beqno
\mathcal{D} F(t,\omega) = \lim_{h \searrow  0}\frac{F(t+h,\omega_t)-F(t,\omega_t)}{h}.
\eeqno
The vertical derivative is defined by $\nabla_\omega F(t,\omega)  = (\partial_iF(t,\omega), i=1,...,n)$, where
\beqno
\partial_iF(t,\omega) = \lim_{h\rightarrow 0} \frac{F(t,\omega_t+  e_ih1_{[t,T]})-F(t,\omega_t)}{h}.
\eeqno
The second order vertical derivative is obtained  by vertically differentiating the elements of vertical derivative, i.e. $\nabla_{\omega}^2F(t,\omega)  = (\partial_j (\partial_iF(t,\omega)) , i,j=1,...,n).$
\brem \label{reduce-std-der} {\normalfont 
If $F(t,\omega)=f(t,\omega(t))$ where $f(t,x)$ is a sufficiently differentiable function $[0,T]\times\mathbb{R}^n\rightarrow\mathbb{R}$, then the horizontal and vertical derivatives reduce to the standard partial derivatives in the sense that $\mathcal{D} F(t,\omega) = \frac{\partial f(t,\omega(t))}{\partial t}$ and $\partial_iF(t,\omega) = \frac{\partial f(t,\omega(t))}{\partial x_i}$. 
}
\erem
From now on we consider a stochastic basis ${\pa{\Omega,{\cal F},\mathbb{P},\underline{\cal F}}}$ where $\underline{\mathcal{F}} = \krull{\mathcal{F}_t}_{0\leq t \leq T}$ is the $\mathbb{P}$-augmented filtration generated by an $n$-dimensional Wiener process $W$. The first main result of functional It\^o calculus is the functional It\^o formula, see e.g. \cite[Theorem 6.2.3]{ContBook} or \cite[Theorem 4.1]{cont2013}.  It can be described as essentially the standard It\^o formula for non-anticipative functionals, where the partial derivatives are replaced with the horizontal and vertical derivatives. The functional It\^o formula holds for non-anticipative functionals which satisfy certain conditions regarding mainly continuity and boundedness written as $ F\in \mathbb{C}_b^{1,2}$, see \cite[Ch.5,6]{ContBook} for details.

The functional It\^o formula implies that if $Y$ is a martingale given by $Y(t) = F(t,W_t)$ $\mathbb{P}$-a.s. for some $F\in \mathbb{C}_b^{1,2}$, then, for any $t$, 
\beqno
Y(t) = Y(0) + \int_0^t\nabla_\omega F(s,W_s)'dW(s)\quad \mathbb{P}\mbox{-a.s.}
\eeqno
In this particular case the vertical derivative of the martingale $Y$ with respect to $W$ is defined by $\nabla_WY(t) = \nabla_\omega F(t,W_t)$.

Let us now extend the definition of the vertical derivative $\nabla_WY$. Let $\mathcal{L}^2(W)$ be the 
space of progressively measurable  processes $\phi$ with 
 $\mathbb{E}[\int_0^T\phi(s)'\phi(s)ds]<\infty$. Let $\mathcal{M}^2(W)$ be the space of square integrable martingales starting at zero.  Let $\mathcal{C}_b^{1,2}(W)$ be the space of processes $Y$ which can be represented as $Y(t) = F(t,W_t)$ $\mathbb{P}$-a.s. with $F\in\mathbb{C}_b^{1,2}$. 

Let $D(W) = \mathcal{C}_b^{1,2}(W)  \cap  \mathcal{M}^2(W)$. It turns out that $\{\nabla_WY \enskip | \enskip  Y \in D(W)\}$ is dense in $\mathcal{L}^2(W)$ and that $D(W)$ is dense in $\mathcal{M}^2(W)$, see \cite[Lemma 7.3.1]{ContBook}. Moreover, the vertical derivative of $Y\in D(W)$, defined above, is characterized as the unique element in $\mathcal{L}^2(W)$ satisfying
\beqno
\mathbb{E}[Y(T)Z(T)] = \mathbb{E}\bracket{\int_0^T\nabla_WY(t)'\nabla_WZ(t)dt} 
\eeqno
for every $Z\in D(W)$, see \cite[Proposition 7.3.2]{ContBook}. Using these observations it is possible to show that the vertical derivative $\nabla_W:D(W) \rightarrow \mathcal{L}^2(W)$ has a unique continuous extension
$\nabla_W:\mathcal{M}^2(W) \rightarrow \mathcal{L}^2(W)$ satisfying
\beqno
\nabla_W\bracket{\int_0^\cdot\phi(s)'dW(s)} = \phi.
\eeqno
Specifically, for $Y \in \mathcal{M}^2(W)$ the (weak) vertical derivative $\nabla_W Y$ is the unique element in $\mathcal{L}^2(W)$ satisfying
\beqno
\mathbb{E}[Y(T)Z(T)] = \mathbb{E}\bracket{\int_0^T\nabla_WY(t)'\nabla_WZ(t)dt}
\eeqno
for every $Z\in D(W)$, see \cite[Theorem 7.3.3]{ContBook}. The martingale representation follows, see e.g. \cite[Theorem 7.3.4.]{ContBook}.
\bth \label{mgthm1} Let $Y$ be a square integrable martingale. Then, for any $t$,
\beqno
Y(t) = Y(0) + \int_0^t \nabla_WY(s)'dW(s) \quad \mathbb{P}\mbox{-a.s.}
\eeqno
\eth
We remark that \cite{lindensjo2016constructive} contains an extension of this result to local martingales.

\section{The financial market} \label{Setup}
This section introduces a standard Wiener driven continuous time financial market that is arbitrage free and complete. For a more detailed description of the market we refer to \cite{Karatzas1} and for proofs that it is arbitrage free and complete we refer to \cite[Ch. 1: Theorem 4.2, Theorem 6.6]{Karatzas1}.

The financial market corresponds to the stochastic basis ${\pa{\Omega,{\cal F},\mathbb{P},\underline{\cal F}}}$ defined in Section \ref{preliminaries}. The market is endowed with  a money market process $B$ defined by 
\beqno
B(t) = e^{\int_0^tr(s)ds}, 0 \leq t \leq T,
\eeqno
where $r$ is a progressively measurable instantaneous risk-free rate process satisfying $\int_0^T|r(t)|dt<\infty$ $\mathbb{P}$-a.s. The market is also endowed with $n$ stocks with price-per-share processes $S_i, i=1,...,n$ which are continuous, strictly positive and satisfy 
\beqno
dS_i(t) = S_i(t)\alpha_i(t)dt + S_i(t)\sum_{d=1}^n\sigma_{id}dW^{(d)}(t), \quad S_i(0)>0, 	\quad 0\leq t \leq T.
\eeqno
It is assumed that the $n$-dimensional process $\alpha$ is progressively measurable and that $\int_0^T|\alpha(t)|dt<\infty$ $\mathbb{P}$-a.s. Moreover, the $n \times n$-dimensional matrix-valued process $\sigma$ is progressively measurable, $\sigma(t)$ is non-singular for all $t$ and all $\omega$ and $\sum_{i=1}^n\sum_{d=1}^n\int_0^T\sigma_{id}^2(t)dt<\infty$ $\mathbb{P}$-a.s. 

The  market price of risk process $\theta$ is defined by 
\beqno
\theta(t) = \sigma(t)^{-1}(\alpha(t)-r(t)\bold{1}), \quad 0 \leq t \leq T.
\eeqno
The  likelihood process $Z$ is defined by
\beqno
Z(t) = e^{-\int_0^t\theta(s)'dW(s)-\frac{1}{2}\int_0^t|\theta(s)|^2ds}, \quad 0 \leq t \leq T.
\eeqno
The  state price density process $H$ is defined by
\beq \label{H-dyn}
H(t) = B(t)^{-1}Z(t)
, \quad 0 \leq t \leq T.
\eeq
\bass \label{ass-Z} $\int_0^T|\theta(t)|^2dt<\infty$ $\mathbb{P}$-a.s. The local martingale $Z$ is a martingale. $\mathbb{E}[H(T)]<\infty$.
\eass

\bdf \label{portf} A  portfolio process $(\pi,\pi^0)$ consists of an $n$-dimensional progressively measurable process $\pi$ and a $1$-dimensional progressively measurable process $\pi^0$ for which $\int_0^T |\pi^0(t) + \pi(t)'\bold{1}||r(t)|dt <\infty$, $\int_0^T |\pi'_t(\alpha(t) -r(t)\bold{1})|dt <\infty$ and $\int_0^T |\pi(t)'\sigma(t)|^2dt <\infty$ $\mathbb{P}$-a.s. The corresponding  wealth process  $X$ is given by
\begin{align} \label{wealth-dyn}
 X(t) & = x_0 + \int_0^t (\pi^0(s) + \pi(s)'\bold{1})r(s)ds  + \int_0^t\pi(t)'(\alpha(s)-r(s)\bold{1})ds  \nonumber \\
& \quad + \int_0^t\pi(s)'\sigma(s)dW(s), \quad 0 \leq t \leq T
\end{align}
where $x_0$ is the initial wealth. The portfolio process is said to be  self-financing if $X(t) = \pi^0(t) + \pi(t)'\bold{1}, 0 \leq t \leq T$. A self-financing portfolio corresponding to $\pi$ is from now on denoted by $X^\pi$.
\edf
Note that the vector $\pi(t)$ corresponds to the amount of capital invested in each stock at the time $t$ and that $B(t)$ corresponds to the amount of capital invested in the money market.

\section{The optimal portfolio process} \label{sec:optimal-port}
For any fixed initial wealth $x_0 \geq 0$, a portfolio process $\pi$ is said to be \emph{admissible} if the corresponding wealth process is self-financing and satisfies $X^\pi(t) \geq 0,0 \leq t \leq T$ $\mathbb{P}$-a.s.

For a fixed initial wealth $x_0 > 0$ we consider the optimal investment problem
\beqno
\sup_{\pi \in \mathcal{A}(x_0)}\mathbb{E}\bracket{U(X^\pi(T))}
\eeqno 
where $\mathcal{A}(x_0)$ is the set of admissible portfolio processes which satisfy the condition $\mathbb{E}[min[U(X^\pi(T),0)]]>-\infty$ and $U$ is a utility function satisfying standard conditions, see \cite[Ch. 3]{Karatzas1}.

Let $I$ denote the (generalized) inverse of the derivative $U'$, for details  see Ch. 3.4 (ibid.). We need the following assumption and standard result. For a proof see Chapter 3 Theorem 7.6, and also Theorem 3.5, Corollary 6.5, Remark 6.4 and p. 102 (ibid.). 
\bass \label{squareINT} 
$\mathbb{E}\bracket{(H(T)I(yH(T)))^2}<\infty, \forall y \in (0, \infty)$.
\eass
\bth \label{optmalXprocess-THM} Consider an initial wealth $x_0 \in (\lim_{y\rightarrow \infty}\mathbb{E}\bracket{H(T)I(yH(T))},\infty)$. The optimal wealth process $X^*$ is then given by 
\beq \label{opt-wealthKandS}
X^*(t) = \mathbb{E}_{\mathcal{F}_t}\bracket{\frac{H(T)}{H(t)}I(\mathcal{Y}(x_0)H(T))}, \quad 0 \leq t \leq T,
\eeq
where $\mathcal{Y}(x_0)>0$ 
is determined by 
\beqno
\mathbb{E}[H(T)I(\mathcal{Y}(x_0)H(T))] = x_0.
\eeqno
\eth
We are now ready to present the main result of the present paper.

\bth \label{mainres1} Consider an initial wealth $x_0 \in (\lim_{y\rightarrow \infty}\mathbb{E}\bracket{H(T)I(yH(T))},\infty)$. The optimal portfolio process $\pi^*$ is then given by
\beq \label{res-x}
\pi^*(t) = 
\sigma(t)'^{-1} \frac{\nabla_{W}\mathbb{E}_{\mathcal{F}_t}\bracket{H(T)I(\mathcal{Y}(x_0)H(T))} +\theta(t)\mathbb{E}_{\mathcal{F}_t}\bracket{H(T)I(\mathcal{Y}(x_0)H(T))}}{H(t)},
\eeq
$0 \leq t \leq T$, where $\nabla_W$ is the vertical derivative operator with respect to $W$.
\eth
\brem {\normalfont Use \cref{opt-wealthKandS} and \cref{res-x} to see that the optimal portfolio process can also be represented as}
\beq \label{res-x2}
\pi^*(t) = \sigma(t)'^{-1} \bracket{H(t)^{-1}\nabla_{W}[H(t)X^*(t)] +\theta(t)X^*(t)}, \quad 0 \leq t \leq T.
\eeq 
\erem
\proof Define $M$ by
\beq \label{pfhelpXX}
M(t)=H(t)X^*(t)= \mathbb{E}_{\mathcal{F}_t}\bracket{H(T)I(\mathcal{Y}(x_0)H(T))]}.
\eeq
Using 
Assumption \ref{squareINT} and $\mathcal{Y}(x_0)>0$ we obtain
\begin{align*}
\mathbb{E}[M(t)^2] 
& = \mathbb{E}\bracket{\mathbb{E}_{\mathcal{F}_t}\bracket{H(T)I(\mathcal{Y}(x_0)H(T))}^2}\\
& \leq \mathbb{E}\bracket{\mathbb{E}_{\mathcal{F}_t}\bracket{(H(T)I(\mathcal{Y}(x_0)H(T)))^2}}\\
& = \mathbb{E}\bracket{(H(T)I(\mathcal{Y}(x_0)H(T)))^2} <\infty.
\end{align*}
It follows that $M$ is a square integrable martingale.

Now use \eqref{H-dyn}, \eqref{wealth-dyn}, the standard It\^o formula, the self-financing condition and the definition of $\theta$ to obtain
\begin{align*}
 dM(t)&=H(t)dX^*(t) + X^*(t)dH(t) + dX^*(t)dH(t)\\
      &= H(t)[X^*(t)r(t)dt + \pi^*(t){'}(\alpha(t)-r(t)\bold{1})dt +  \pi^*(t){'}\sigma(t)dW(t)] \\
			& \quad + X^*(t)[-r(t)H(t)dt -\theta(t)'H(t)dW(t)]\\
			& \quad + \pi^*(t){'}\sigma(t)(-\theta(t)H(t))dt\\
			& = H(t)\pi^*(t){'}\sigma(t)dW(t) -X^*(t)\theta(t)'H(t)dW(t).
\end{align*}
This implies that $\pi^*$ satisfies, for any $t$, 
\beqno
M(t) = M(0) + \int_0^tH(s)(\pi^*{'}(s)\sigma(s)-X^*(s)\theta(s)')dW(s).
\eeqno
Since $M$ is a square integrable martingale we may  use Theorem \ref{mgthm1} to obtain, for any $t$, the representation
\beqno
M(t) = M(0) + \int_0^t\nabla_{W}M(s)'dW(s) \quad \mathbb{P}\mbox{-a.s.}
\eeqno
It follows that $\pi^*$ can be represented by
\beqno 
\nabla_{W}M(t)' = H(t)(\pi^*{'}(t)\sigma(t)-X^*(t)\theta(t)').
\eeqno
It follows that
\beqno
\pi^*(t)' \sigma(t)= \bracket{\frac{\nabla_{W}M(t)'}{H(t)}  + X^*(s)\theta(t)'},
\eeqno
which implies that
\beq \label{pflast}
\pi^*(t)= \sigma(t)'^{-1} \bracket{\frac{\nabla_{W}M(t)}{H(t)}  + X^*(t)\theta(t)}.
\eeq
Replace $X^*(t)$ in \cref{pflast} with the right side of \cref{opt-wealthKandS} and replace $M(t)$ with the right side of \cref{pfhelpXX}. The result follows.
\endproof

\subsection{A comparison with the Malliavin calculus approach} \label{mall-comp}
According to the Clark-Ocone theorem it holds that if $F$ is an integrable $\mathcal{F}_T$-measurable random variable that is Malliavin differentiable in the sense $F \in \textbf{D}_{1,1}$ then 
\begin{align*}
F = \mathbb{E}[F] + \int_0^T\mathbb{E}_{\mathcal{F}_t}[(D_tF)']dW(t),
\end{align*}
where $D$ is the Malliavin derivative operator. For a definition of the space $\textbf{D}_{1,1}$ and a proof we refer to \cite[Appendix E]{Karatzas1} and \cite{karatzas1991extension}.

With the Clark-Ocone theorem as a starting point \cite{oconeKaratzas} arrives at an explicit formula for the optimal portfolio process $\pi^*$ based on Malliavin derivatives, in essentially the same financial market that we study in the present present paper. Naturally, this result relies on the requirement that the discounted optimal terminal wealth is sufficiently Malliavin differentiable. In order to ensure that this condition is fulfilled, further restrictions on the financial market and the utility function are needed. For example, conditions regarding the Malliavin differentiability of $\theta$ and $r$, and further conditions for the inverse $I$ are necessary, see \cite[Theorem 4.2]{oconeKaratzas} for details.

In comparison, the only non-standard condition that the explicit formula for $\pi^*$ in Theorem \ref{mainres1} relies on is the square integrability in Assumption \ref{squareINT}, since typically only integrability is assumed, cf. \cite[Ch. 3.7]{Karatzas1}.

\subsection{Examples} \label{examples}
Let us illustrate Theorem \ref{mainres1} by studying two well-known examples.
\subsubsection{Logarithmic utility}
Let $U(x) = ln(x)$ for $x\in (0,\infty)$. It follows that $I(y) = \frac{1}{y}$ for $y \in(0,\infty)$. 
Using calculations similar to those in \cref{log-ex1} below it is easy to see that Assumption \ref{squareINT} is satisfied and that $\lim_{y\rightarrow \infty}\mathbb{E}\bracket{H(T)I(yH(T))} = 0$. We may therefore use Theorem \ref{optmalXprocess-THM} and Theorem \ref{mainres1} for any initial wealth $x_0>0$. We directly obtain 
\beq \label{log-ex1}
\mathbb{E}_{\mathcal{F}_t}\bracket{H(T)I(\mathcal{Y}(x_0)H(T))} =  \mathbb{E}_{\mathcal{F}_t}\bracket{H(T)\frac{1}{\mathcal{Y}(x_0)H(T)}} = 
\frac{1}{\mathcal{Y}(x_0)}.
\eeq
 This implies that
\beq \label{log-ex12z}
\nabla_{W}\mathbb{E}_{\mathcal{F}_t}\bracket{H(T)I(\mathcal{Y}(x_0)H(T))} = \nabla_{W}\bracket{\frac{1}{\mathcal{Y}(x_0)}} = 0,
\eeq
since the vertical derivative reduces to the standard derivative in this case, cf. Remark \ref{reduce-std-der}. Using \cref{res-x}, \cref{log-ex1} and \cref{log-ex12z} we obtain
\beqno
\pi^*(t) = 
\sigma(t)'^{-1} \frac{0 + \theta(t)\frac{1}{\mathcal{Y}(x_0)}}{H(t)} = \frac{(\sigma(t)\sigma(t)')^{-1}(\alpha(t)-r(t)\bold{1})}{\mathcal{Y}(x_0)H(t)}.
\eeqno
Now use Theorem \ref{optmalXprocess-THM} and \cref{log-ex1} to see that $x_0 = \frac{1}{\mathcal{Y}(x_0)}$ which implies that the optimal portfolio can be represented as
\beqno
\pi^*(t) = (\sigma(t)\sigma(t)')^{-1}(\alpha(t)-r(t)\bold{1})\frac{x_0}{H(t)}.
\eeqno
For completeness sake we use Theorem \ref{optmalXprocess-THM} and the above to obtain the optimal wealth process
\beqno
X^*(t) 
= \frac{x_0}{H(t)}.
\eeqno

\subsubsection{Power utility with deterministic coefficients}
Let $r,\alpha$ and $\sigma$ be deterministic functions of time and $U(x)= \frac{x^\gamma}{\gamma}$ for $x\in (0,\infty)$ with $\gamma<1,\gamma\neq 0$. It follows that $I(y) = y^{\frac{1}{\gamma-1}}$ for $y\in (0,\infty)$. This implies that
\beqno
\mathbb{E}\bracket{(H(T)I(yH(T)))} = y^{\frac{1}{\gamma-1}} \mathbb{E}\bracket{H(T)H(T)^{\frac{1}{\gamma-1}}} =
y^{\frac{1}{\gamma-1}} \mathbb{E}\bracket{H(T)I(H(T))}.
\eeqno
Assumption \ref{squareINT} implies that $\lim_{y\rightarrow \infty}\mathbb{E}\bracket{H(T)I(yH(T))} = 0$. A sufficient condition for Assumption \ref{squareINT} is in this case that $\theta$ and $r$ are bounded.

Recall that $HX^*$ given by
\beqno
H(t)X^*(t) = \mathbb{E}_{{\mathcal{F}_t}}\bracket{H(T)I(\mathcal{Y}(x_0)H(T))},
\eeqno
is a square integrable martingale. Now use $I(y) = y^{\frac{1}{\gamma-1}}$, \cref{H-dyn} and \cref{opt-wealthKandS} to perform the following calculations, where $(...)$ denotes a deterministic function of time based on $r,\theta$ and $\gamma$,
\begin{align}
H(t)X^*(t) 
& = \mathcal{Y}(x_0)^{\frac{1}{\gamma-1}} \mathbb{E}_{{\mathcal{F}_t}}\bracket{H(T)^{\frac{\gamma}{\gamma-1}}}   \label{pow1}\\
& = \mathcal{Y}(x_0)^{\frac{1}{\gamma-1}} \mathbb{E}_{{\mathcal{F}_t}}\bracket{e^{-\int_0^T\frac{\gamma}{\gamma-1}\theta(s)'dW(s) + \int_0^T(...)ds}} \nonumber\\
& = \mathcal{Y}(x_0)^{\frac{1}{\gamma-1}} e^{\int_0^T(...)ds}\mathbb{E}_{{\mathcal{F}_t}}\bracket{e^{\int_0^T\frac{-\gamma}{\gamma-1}\theta(s)'dW(s)}}\nonumber\\
& = \mathcal{Y}(x_0)^{\frac{1}{\gamma-1}} e^{\int_0^T(...)ds}e^{\int_0^t\frac{-\gamma}{\gamma-1}\theta(s)'dW(s)} \mathbb{E}_{{\mathcal{F}_t}}\bracket{e^{\int_t^T\frac{-\gamma}{\gamma-1}\theta(s)'dW(s)}}\nonumber\\
& = \mathcal{Y}(x_0)^{\frac{1}{\gamma-1}} e^{\int_0^T(...)ds}e^{\int_0^t\frac{-\gamma}{\gamma-1}\theta(s)'dW(s)} e^{\frac{1}{2}\int_t^T|\frac{\gamma \theta(s)}{\gamma-1}|^2ds}\nonumber\\ 
& = \mathcal{Y}(x_0)^{\frac{1}{\gamma-1}}e^{\int_0^T(...)ds}e^{\int_0^t\frac{-\gamma}{\gamma-1}\theta(s)'dW(s)}
e^{\frac{1}{2}\int_0^T|\frac{\gamma \theta(s)}{\gamma-1}|^2ds-\frac{1}{2}\int_0^t|\frac{\gamma\theta(s)}{\gamma-1}|^2ds}\nonumber\\
& = \mathcal{Y}(x_0)^{\frac{1}{\gamma-1}}e^{\int_0^T(...)ds}e^{\int_0^t\frac{-\gamma}{\gamma-1}\theta(s)'dW(s)}
e^{-\frac{1}{2}\int_0^t|\frac{\gamma\theta(s)}{\gamma-1}|^2ds}\nonumber.
\end{align}
Thus, $HX^*$ is in fact a square integrable exponential martingale. Using It\^o's formula we obtain
\beqno
H(t)X^*(t) = H(0)X^*(0) + \int_0^tH(s)X^*(s)\frac{-\gamma}{\gamma-1}\theta(s)'dW(s).
\eeqno
Together with Theorem \ref{mgthm1} this implies that the vertical derivative of $HX^*$ with respect to $W$ can be represented as
\beq \label{powereq11}
\nabla_W[H(t)X^*(t)]' = H(t)X^*(t)\frac{-\gamma}{\gamma-1}\theta(t)'.
\eeq 
Now use  \cref{res-x2} and \cref{powereq11}  to see that the optimal portfolio can be represented as
\begin{align*}
\pi^*(t) &  = \sigma(t)'^{-1} \bracket{H(t)^{-1}\nabla_{W}[H(t)X^*(t)] +\theta(t)X^*(t)}\\
& = \sigma(t)'^{-1} \bracket{X^*(t)\frac{-\gamma}{\gamma-1}\theta(t) +\theta(t)X^*(t)}\\
& = \sigma(t)'^{-1} \bracket{\frac{1}{1-\gamma}\theta(t)X^*(t)}\\
& = (\sigma(t)\sigma(t)')^{-1} (\alpha(t)-r(t)\bold{1})\frac{X^*(t)}{1-\gamma}.
\end{align*}
For completeness sake let us find an expression for the optimal wealth process $X^*$. Use Theorem \ref{optmalXprocess-THM} and $I(y) = y^{\frac{1}{\gamma-1}}$ to see that
\beqno
\mathcal{Y}(x_0)^{\frac{1}{\gamma-1}} = \frac{x_0}{\mathbb{E}\bracket{H(T)^{\frac{\gamma}{\gamma-1}}}}.
\eeqno
Using \cref{pow1} and the above we obtain
\beqarno
X^*(t) =  \frac{1}{H(t)}\frac{x_0}{\mathbb{E}\bracket{H(T)^{\frac{\gamma}{\gamma-1}}}}\mathbb{E}_{{\mathcal{F}_t}}\bracket{H(T)^{\frac{\gamma}{\gamma-1}}}.
\eeqarno

\bibliography{kristofferBibl}

\end{document}